\documentclass[aps,prl,reprint,amssymb,longbibliography,preprintnumbers,twocolumn,amsmath,superscriptaddress,floatfix]{revtex4-2}
\usepackage{graphicx}
\usepackage{bm}
\usepackage{hyperref}
\usepackage{physics}
\usepackage[final]{showlabels}
 \usepackage{color}
\usepackage[normalem]{ulem}

\begin{document}

\title{High-harmonic spectroscopy of strongly bound excitons in solids}

\author{Simon Vendelbo Bylling Jensen}
\affiliation{Department of Physics and Astronomy, Aarhus University, DK-8000 Aarhus C, Denmark}
    
\author{Lars Bojer Madsen}
\affiliation{Department of Physics and Astronomy, Aarhus University, DK-8000 Aarhus C, Denmark}

\author{Angel Rubio}
\email{angel.rubio@mpsd.mpg.de}
\affiliation{Max Planck Institute for the Structure and Dynamics of Matter and Center for Free-Electron Laser Science, Hamburg 22761, Germany}
\affiliation{Center for Computational Quantum Physics (CCQ), The Flatiron Institute, New York, New York 10010, USA}

\author{Nicolas Tancogne-Dejean}
\email{nicolas.tancogne-dejean@mpsd.mpg.de}
\affiliation{Max Planck Institute for the Structure and Dynamics of Matter and Center for Free-Electron Laser Science, Hamburg 22761, Germany}

\date{\today}
\begin{abstract}
We explore the nonlinear response of ultrafast strong-field driven excitons in a one-dimensional solid with \textit{ab initio} simulations. We demonstrate from our simulations and analytical model that a finite population of excitons imprints unique signatures to the high-harmonic spectra of materials. We show the exciton population can be retrieved from the spectra. We further demonstrate signatures of exciton recombination and that a shift of the exciton level is imprinted into the harmonic signal. The results open the door to high-harmonic spectroscopy of excitons in condensed-matter systems. %

\end{abstract}  
\maketitle

When semiconductor or insulator systems interact with light, their optical linear and non linear responses is often dominated by features arising due to bound electron-hole pairs known as excitons \cite{RevModPhys.83.543}. 
Excitons are consequential in photonic technology \cite{PhysRevB.94.041401} and play a significant role in many processes, such as energy transfer and light absorption in bio-molecules~\cite{bardeen2014structure, spano2006excitons}, nanostructures~\cite{spano2006excitons}, and solids\cite{cudazzo2015exciton,mueller2018exciton}, and have fundamental and technological applications.
Bound excitons give rise to sharp peaks in absorption and photoluminescence spectra, which exhibit optical features such as the Stark effect \cite{PhysRevLett.92.157401}, Rabi oscillations \cite{PhysRevLett.87.133603} and Fano resonances \cite{PhysRevLett.74.470}. 
Understanding how they behave under external perturbations such as ultrafast intense fields is therefore key to make use of them for future applications like PHz electronics and guide experimental observations by providing microscopical understanding of their pump probe date.

Excitons arise in different forms, such as interlayer excitons, optically dark excitons, strongly bound excitons  to name a few \cite{Mueller2018,Merkl2019}, and are important for the properties of bulk materials but also low-dimensional materials~\cite{RevModPhys.90.021001,qiu2019giant,manzeli20172d,PhysRevLett.113.026803}, van der Walls heterostructures \cite{Rivera2015}, and twistronics ~\cite{patel2015tunable,Tran2019}. 
The development of methods allowing to study the dynamics of excitons, is an active area of research, including the study of exciton formation~\cite{trovatello2020ultrafast}, ultrafast dynamics ~\cite{pogna2016photo,jiang2021real,PhysRevLett.128.016801}, condensation~\cite{PhysRevB.103.L241404}, dissociation~\cite{massicotte2018dissociation}, and coupling to other degrees of freedom~\cite{PhysRevB.105.085111,li2021exciton}. 

While excitons are know to play important roles in many aspects of material science, and can even dominate in linear and perturbative nonlinear spectroscopies in solids, it is common to neglect excitonic effects in describing electron dynamics induced by intense laser fields. This approach is in the spirit of the strong-field approximation of atomic physics, in which the laser field 
is assumed to dominate over the Coulomb interaction~\cite{PhysRevA.49.2117} , thus motivated a description in terms of independent particles. 
It is thus rationalized, that either excitons do not form or that any bound exciton present in the material would dissociate during strong-field processes \cite{RevModPhys.90.021002}. In this work, however, we show that for typical laser parameters used for strong-field physics in solids, this argument fails for strongly bound excitons. Indeed, the latter are shown to modify the ultrafast optical response of condensed matter systems. 

With recent experiments on nonlinear exciton dynamics in THz harmonic sideband generation  \cite{Langer2016,Zaks2012} and in attosecond transient reflection and absorption spectroscopy \cite{Lucchini2021,PhysRevLett.124.207401,doi:10.1126/science.aan4737,Kobayashi2023}, it is crucial to elucidate the dynamics of excitons under intense laser fields, in order to support a complete understanding of light-matter interactions. Here we consider their impact on the process of high-order harmonic generation (HHG). HHG utilizes the ultrafast nonlinear response of a material to generate ultrashort coherent pulses, which inherit spectrographic information from the underlying electron dynamics \cite{PhysRevA.66.023805,PhysRevLett.98.203007,doi:10.1126/science.1163077,doi:10.1126/science.1123904,PhysRevLett.94.053004,Itatani2004,Schubert2014,Garg2018,PhysRevLett.115.193603}. 
So far, most condensed-matter HHG experiments are rationalized in terms of independent-electron models, which ignore excitons, but are capable of describing energy-cutoff scaling, spectral features \cite{Ghimire2011,Schubert2014,Luu2015,PhysRevB.102.104308,heide2021probing}, orientation and polarization dependencies \cite{You2017,PhysRevLett.120.243903}, as well as reconstruction of bandstructure \cite{Lanin:17,Lanin:19} and Berry curvature \cite{Luu2018} even if topological signatures in HHG remain elusive ~\cite{2303.17300}. Experimental indications of possible excitonic effects have arisen in HHG as a plateau %
in rare-gas solids \cite{Ndabashimiye2016}, an increased efficiency of monolayer compared to bulk crystals \cite{Liu2017} and a characteristic delay-dependency in pump-probe HHG \cite{heide2021probing}. However, a clear demonstration of excitonic effects related to a controlled exciton population in HHG remains elusive.
In the semiconductor Bloch equations (SBE) formalism, it was already indicated that excitons could be important for interpreting the HHG mechanisms, as the relative importance of inter- or intraband contributions are altered if the Coulomb interaction is tuned to reproduce accurate exciton binding energies %
\cite{Garg2016}. Recently, excitons have also been predicted to influence HHG in Mott insulators and monolayer transition metal dichalcogenides, in the framework of effective Hamiltonian models \cite{PhysRevB.105.L241108, PhysRevResearch.2.023072,PhysRevA.105.063504}. 

A few approaches may capture excitons in real-time \textit{ab initio} simulations, like non-equilibrium Green's functions (NEGF) ~\cite{PhysRevB.37.941, stefanucci2013nonequilibrium,balzer2012nonequilibrium}, based on the generalized Kadanoff-Baym ansatz ~\cite{PhysRevB.34.6933}. While important progresses have been made ~\cite{PhysRevLett.128.016801,PhysRevB.105.125134}, this method is still numerically prohibitive, and one needs to employ simpler methods like time-dependent Hartree-Fock (TDHF)\cite{PhysRevLett.33.582}, or the related hybrid functionals in TDDFT~\cite{PhysRevLett.89.096402}.
Hybrid functionals allow to explore ultrafast and nonlinear electrons dynamics \cite{PhysRevLett.127.077401} with low-cost and accurate alternatives to NEGF~\cite{PhysRevResearch.2.013091} but is restricted on e.g. the dimensionality. Furthermore, the formalism within TDDFT to provide access to time-resolved visualization of the exciton wavefunction \cite{doi:10.1021/acs.jctc.0c01334}, and thus provide insights on the exciton dynamics in space and time is also applicable for the more established TDHF formalism.

In light of this, we perform here real-time \textit{ab initio} wavefunction-based TDHF simulations to characterize nonlinear ultrafast exciton dynamics and how excitons modify the HHG response in realistic pump-probe setting. Simulations are performed for an one-dimensional insulating hydrogen crystal, that has strongly bound excitons ~\cite{PhysRevA.98.023415,PhysRevLett.127.077401}, and is therefore ideal for unraveling the fundamental ultrafast exciton dynamics.
\nocite{doi:10.1063/1.5142502,Lin2016,PhysRevB.75.205126,PhysRevB.62.7998,LI2011157,C5CP07077E,PhysRevB.43.14325,PhysRevB.44.8138,de2015modeling,PhysRevB.91.064302,PhysRevA.102.033105,PhysRevResearch.2.033333,PhysRevA.99.013435} 
Detail are given in Supplemental Material (SM) Sec. I \cite{SM}. 

\begin{figure} 
\includegraphics[width=8.6 cm]{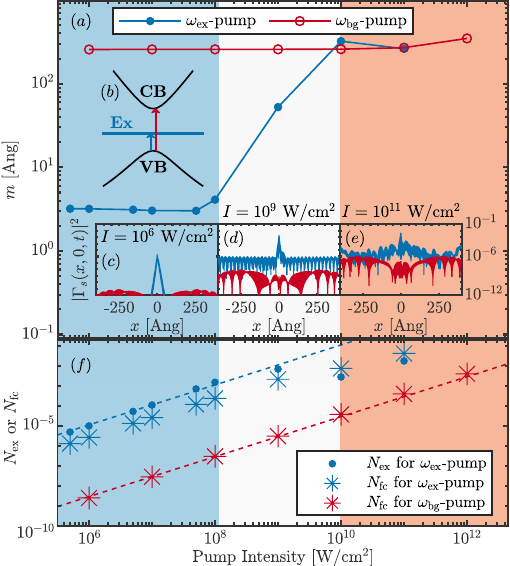}
\caption{(a) First moment, $m$ of Eq.~\eqref{eq:1}, of the approximated exciton wavefunction for different pump frequencies and intensities. The areas denote regimes where the $\omega_{\mathrm{ex}}$-pump generates dominantly bound excitons (blue) or free carriers (red). (b) Excitation pathways from the valence band (VB) to conduction band (CB) or to exciton (Ex). The corresponding energies are $\omega_{\mathrm{bg}} = 9.45$ eV and $\omega_\text{ex}= 3.86$ eV. (c)-(e) Exciton density for a hole at $x'=0$, after excitation by a $25$ fs pump pulse with intensities of $10^6$ W cm$^{-2}$, $10^9$ W cm$^{-2}$, and $10^{11}$ W cm$^{-2}$, respectively. (f) Number of pumped excitons $N_\text{ex}$ and free carriers $N_\text{fc}$ in the system. The dashed lines show the linear perturbative scaling behavior for resonant excitation. The highest intensity value for the $\omega_\mathrm{ex}$-pump is omitted due to the excitation exceeding the damage threshold predicted under the electron-hole plasma model of $\approx$10\% of excited electrons~\cite{PhysRevB.46.10686,PhysRevB.49.7299,PhysRevB.42.7163}. 
The quantities of (a)-(f) are evaluated after the pump preparation, just before the system is driven to produce HHG. (See SM \cite{SM} for pulse and system details.) } \label{fig:1}
\end{figure}

We first investigate how an exciton population in the material is created. For this, we pump the material with a laser under different excitation conditions. We need to define a criterion to isolate the effect of the pump laser regarding excitation of free-carriers or excitons. We determine the nature of the excitation by analyzing the real-time dynamics of the exciton wavefunction~\cite{doi:10.1021/acs.jctc.0c01334} %
From the norm of the approximated exciton wavefunction, $\Gamma_s(x,x',t)$, %
we obtain the conditional probability for an electron to be at the position $x$ while having a hole at $x'$ (see SM Sec. I \cite{SM}). 
The first moment of the exciton wavefunction~\cite{man2021experimental, dong2021direct}
\begin{equation}
 m = \left. \int dx \abs{x} \abs{\Gamma_s (x,0,t)}^2 \middle/  \int dx \abs{\Gamma_s (x,0,t)}^2 \right. \,, \label{eq:1}
\end{equation}
reveals the degree of localization of an excitation (in this case around $x' = 0$), and therefore  can in principle allow us to  distinguish between bound excitons and free carriers. Indeed, if the material contains strongly-bound excitons, $m$ is small since the electron is very likely to remain near its hole. Note that $m$ is used to define the exciton radius from \textit{ab initio} simulation \cite{Prete2020}. Inversely, if a material contains free carriers and conduction bands are dispersive, then electrons are delocalized throughout the crystal, and $m$ increases to the size of the crystal. 
Figure~\ref{fig:1} (a) reveals that while for band-gap resonant pumping we only generate free carriers, for exciton resonant pumping, $m$ attains smaller values suggesting appreciable population of excitons, as expected for resonant pumping~\cite{PhysRevB.37.941}. 
At low intensity, indicated by the blue area, we generate mostly bound excitons, as also shown by the exciton density [Fig.~\ref{fig:1}(c)]. For high intensity, indicated by the red area, the population of free-carriers clearly dominates the excitation. This is also visible from the exciton density [Figs.~\ref{fig:1}(d,e)].
Therefore we use $m$ to differentiate the nature of the excitation, and extract the number of excitons $N_\text{ex}$ and number of free carriers $N_\text{fc}$ for the various excited systems in Fig.~\ref{fig:1}~(f), see SM Sec.~I~\cite{SM}. 
We observe that the generation of excitons by the $\omega_\text{ex}$-pump deviates from the first-order perturbative response, indicated by the dashed blue line. This deviation can be attributed to a subsequent exciton dissociation process occurring during the pumping, as the pump ionize the excitons it creates. This is confirmed by the ionization for a bound exciton model within the effective mass approximation, see SM Sec.~II \cite{SM}. Alternative mechanisms, such as nonlinear effects associated with three-photon excitations directly to the bandgap are excluded as this would reflect in a change of the slope of the power-law in Fig.~\ref{fig:1}~(f). On the contrary, the $\omega_\text{bg}$-pumped systems prepare dominantly free carriers at all intensities and follow the perturbative scaling. %
We note that the difference in the excitation magnitudes for the two pump frequencies in Figs.~\ref{fig:1}(c)-(f) can be rationalized from the relative resonance magnitudes in the linear absorption spectra, see SM Fig.~2~\cite{SM}. 
 In a pump-probe setting, the pump intensity for the exciton transition needs to be selected with care, to not dissociate them with the pump itself.

\begin{figure} 
\includegraphics[width=8.6 cm]{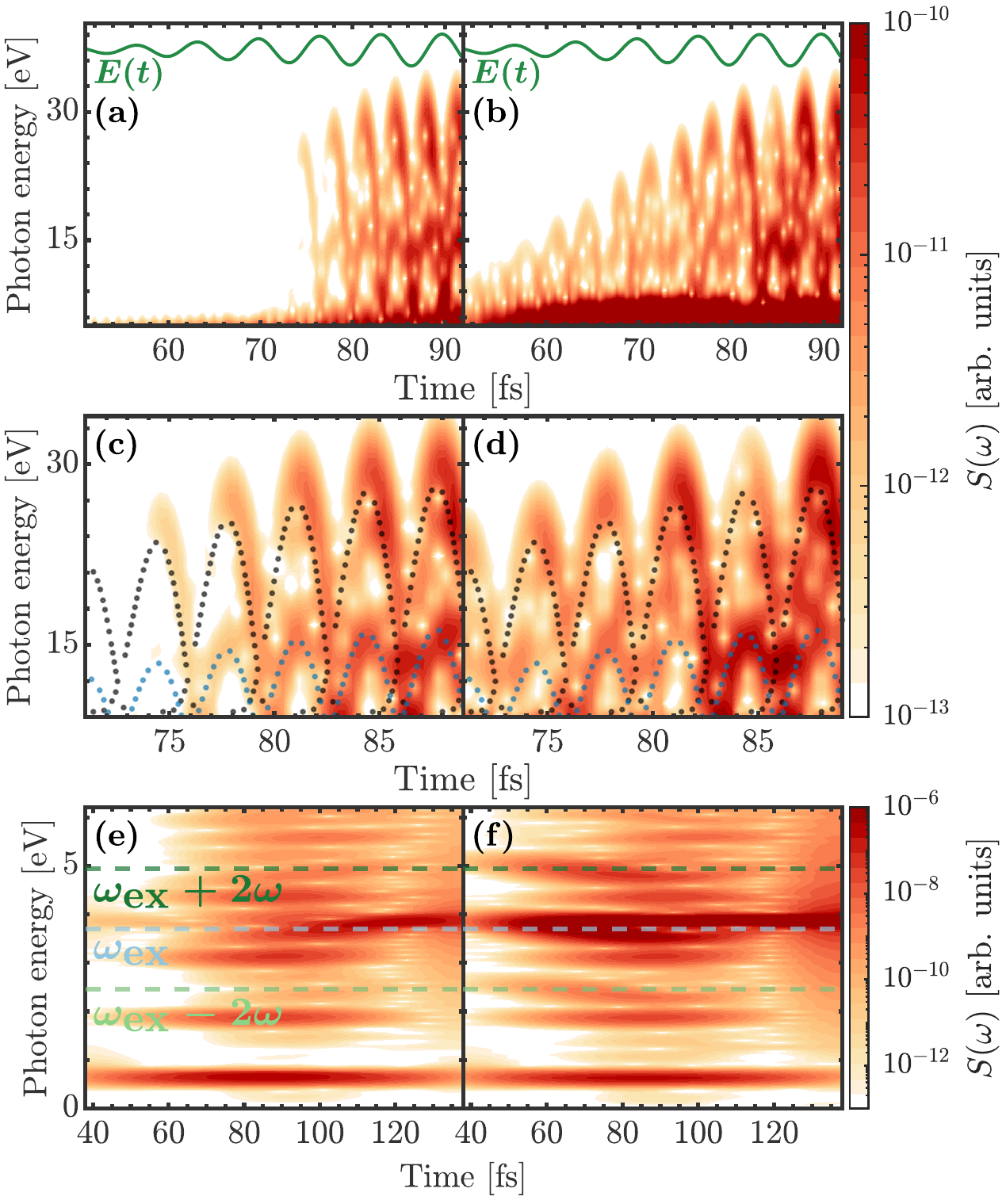}
\caption{
Time-frequency analysis of the harmonic radiation of the region above the exciton binding energy in (a),(b), magnified in (c),(d) and below the bandgap (e),(f) for, respectively, an unpumped (a),(c),(e) and an exciton-seeded sample (b),(d),(f), obtained by an exciton-resonant pump with an intensity of $10^7$ W cm$^{-2}$. The $2000$ nm driving electric field is sketched in green in (a),(d). In (a),(d) we used a window of $\sigma = 0.40$ fs for the Gabor transform, and of width $\sigma = 10$ fs for (e),(f). Trajectories from our exciton-extended semiclassical model, see SM Sec.~V~\cite{SM}, is depicted with dotted lines in (c),(d). Black color depicts trajectories recombining to the valence band and the blue color corresponds to recombination in form of a bound exciton. Dashed lines in (e),(f) denote the locations of exciton peak, as well as the first exciton sidebands.} \label{fig:3}
\end{figure}
We now investigate the effect of a finite exciton population on the HHG spectra, by performing simulations where the material is excited by an exciton-resonant pump compared to the unpumped material. Differences between the considered cases (unpumped, band-gap resonant pumping and exciton-resonant pumping) start to emerge when the pump excites more than $10^{-5}$ valence electrons per unit cell, see SM Sec.~III~\cite{SM}.
To investigate the signatures of exciton dynamics in HHG, we perform a time-frequency analysis, see Fig.~\ref{fig:3} of the harmonic emission %
comparing the exciton-resonant pump case, to the unpumped case. 
Above the bandgap [Figs.~\ref{fig:3} (a)-(b)], the time-frequency analysis reveals that the exciton-seeded system is starting to emit harmonics prior to the unpumped sample. The initial emission time, at $75$ fs in Fig.~\ref{fig:3} (a) correspond to the time where substantial exciton dissociation is observed, see SM Fig.~6~\cite{SM}, indicating that ionization of excitons are primarily responsible for the early stage of the harmonic emission.
Indeed, the dissociate of exciton requires a weaker field strength than the excitation of electrons through the bandgap given the large binding energy of our system. The two-step process in which an exciton is generated and afterwards dissociated could thus play a dominant role.
Based on our results, we note that pumping excitons in the sample looks favorable with regard to decreasing the nonlinear response time of a condensed-matter system compared to an unpumped system. Similarly, pumping excitons in the sample also looks favorable for enhancing interband emission, see  SM Sec.~IV~\cite{SM}.

Commonly, the time-frequency analysis of the harmonic emission in solids consists of two contributions, a chirped emission related to the interband emission channel, and a chirpless emission due to the intraband motion of the electrons~\cite{PhysRevB.107.054302}. However, we observe not one set of chirped trajectories, but two, as shown more clearly in Fig.~\ref{fig:3}~(c)-(d). The standard approach to explain the chirped trajectories is to use a semiclassical interband model that propagates the relative electron-hole distance $\Delta x(t)$ as predicted by the band dispersion $\varepsilon_{j}\left(k(t')\right)$ of the valence and conduction bands ($j\in  \{v,c \}$) with 
\begin{equation}
\Delta x(t) = \int_{t_0}^t \pdv{\left[\varepsilon_{c}\left(k(t')\right) - \varepsilon_{v}\left(k(t') \right) \right]}{k} dt'\,,\nonumber
\end{equation}
where $t_0$ plays the role of the ionization time, see SM Sec.~V~\cite{SM}.
The recombination time $t_r$ is then conditioned by $\Delta x(t_r) = 0$ (i.e. neglecting imperfect recombination discussed in Ref.~\cite{PhysRevLett.124.153204,PhysRevA.95.043416,PhysRevA.102.033105}). This leads to an emission of light at the separation energy $\varepsilon_{c}\left[k\left(t_r\right)\right] - \varepsilon_{v}\left[k\left(t_r\right)\right]$. Using this model, we find the emission pattern shown by the black dots in Fig.~\ref{fig:3} (c)-(d). \\

Here we extended this semiclassical model in order to explain the other set of trajectories.
More precisely, we modify the recombination step to allow recombination into an exciton with binding energy $E_b$ with emission of $\varepsilon_{c}\left[k\left(t_r\right)\right] - E_b$. This leads to the emission pattern shown by the blue dots.
This recombination channel captures the second set of trajectories, thus revealing the importance of exciton recombination in the subcycle nonlinear electron dynamics. Since the energy goes into the formation of the exciton, the resulting harmonic emission energy is reduced, and the trajectories do not affect the high-energy part of the spectrum nor the energy cutoff.
We note that dissociation of excitons can also impact the ionization step of the semiclassical model. However, this leads to the same trajectories as the formation of free carriers when pumping electrons directly to the CBs. We cannot therefore distinguish signatures of exciton dissociation directly from the trajectories or energy cutoff. 
Apart from the new set of exciton-related trajectories, our simulations also reveal the appearance of exciton-related features in the below-bandgap energies, as shown in Figs.~\ref{fig:3} (e) and (f). 
For the unpumped system [Fig.~\ref{fig:3} (e)], a continuous emission of clean odd-order harmonics is observed, %
as well as a resonance corresponding to the energy for generating or annihilating an exciton, as also observed in Ref.~\cite{PhysRevA.98.023415}. This is the expected subcycle dynamics for intraband emission~\cite{PhysRevLett.118.087403}. For the exciton-seeded system [Fig.~\ref{fig:3} (f)] we see that the exciton resonance is enhanced throughout and leads to continuous emission of weaker spectral features, corresponding to sidebands of the exciton resonance, located at $\omega_\mathrm{\mathrm{ex}} \pm 2 \omega$. %
Such sidebands occur as a consequence of the prepared population of excitons being driven by the probe pulse to annihilate. In doing so, the partly dissociated excitons can undergo a laser-assisted electron-hole recollision process where photons can be exchanged with the strong electromagnetic field \cite{Langer2016,Zaks2012}. Opposed to these THz excitonic sideband experiments, here we predict sideband generation, where energies above $1$\,eV are exchanged with the electromagnetic field. For these quasiparticle collisions to occur, the sample needs a significant population of excitons. We note from that the exciton resonance and sidebands are shifted slightly in energy, with increasing probe intensity, see SM Sec.~VI~\cite{SM}.

\begin{figure} 
\includegraphics[width=8.6 cm]{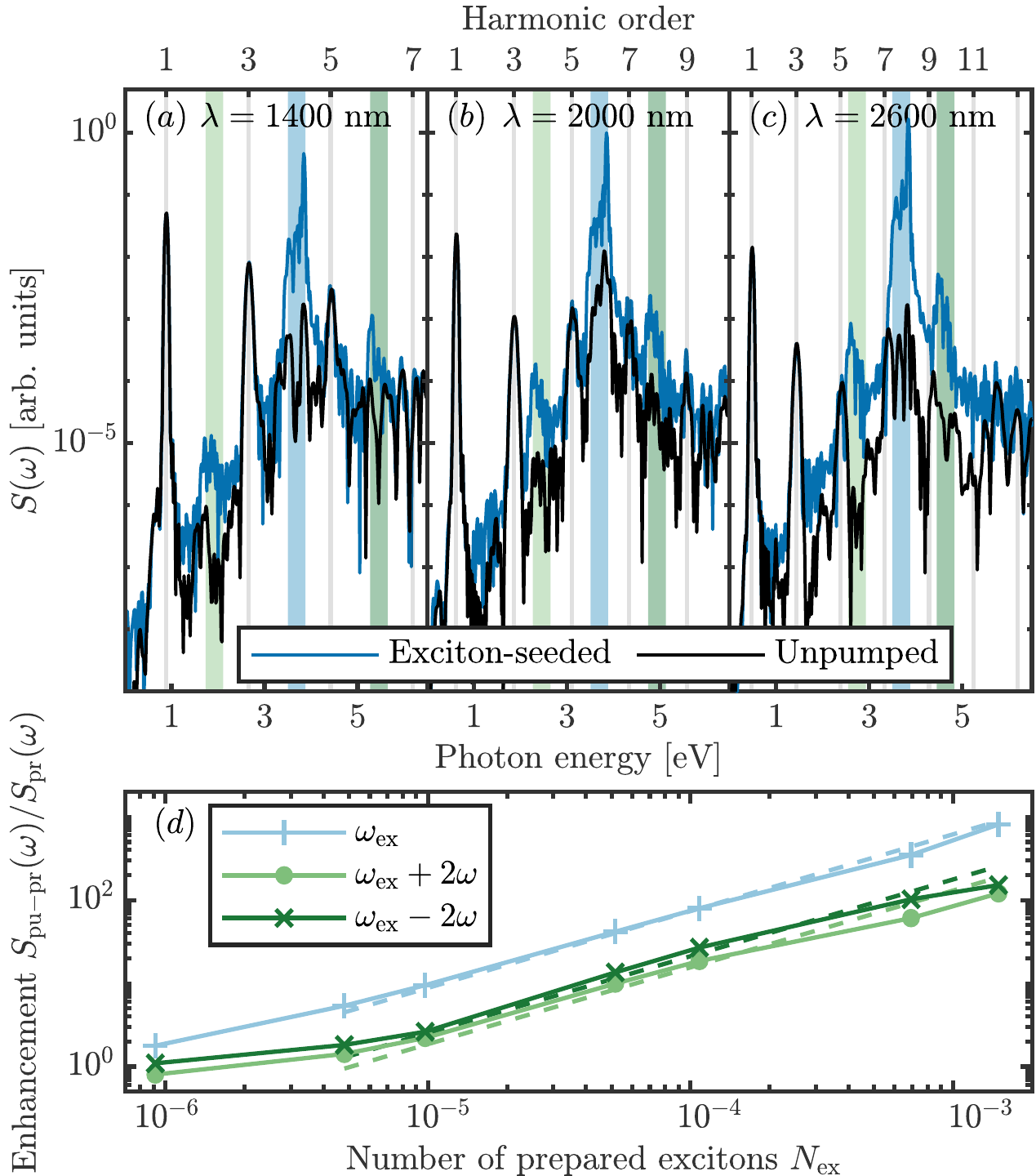}
\caption{ (a)-(c) HHG spectra for various wavelengths for the unpumped system versus the system prepared by an exciton-resonant pump. The HHG driving probe wavelength is scanned across $1600$, $2000$, and $2600$ nm, respectively, for (a)-(c). The colored areas denote the exciton resonance (blue) and the first excitonic associated sidebands, at $\omega_\mathrm{ex} \pm 2\omega$ (green). (d) Harmonic yield enhancement of the exciton resonance and sidebands for $\lambda=2000$ nm as a function of  bound exciton population, utilizing a $10^5 - 10^8$ W cm$^{-2}$ exciton-resonant pump. Dashed lines are explained in the main text.} \label{fig:2}
\end{figure}
To utilize the spectroscopic capabilities of the harmonic exciton resonance and sidebands, we consider the harmonic emission spectra in Figs.~\ref{fig:2} (a)-(c) for a system driven with a $1600$, $2000$ or $2600$ nm probe pulse using an intensity of $10^7$ W cm$^{-2}$ and probe duration of 100\,fs.
All our simulations showed exciton-induced sidebands, irrespective of probe duration and intensity.
The signals from the unexcited system are given by the black curves and provide clean harmonic peaks with the inclusion of an exciton resonance at $\omega_{\mathrm{ex}}$. For the exciton-seeded system, the bound-exciton population contributes to an enhancement of the harmonic spectrum at the exciton energy, and its spectral sidebands in the regions marked with blue and green. This result shows that the sidebands can be observed over a laser range of wavelengths and are a robust feature of HHG from exciton-pump materials. %
Using our simulated pump-probe set-up, we now vary the exciton population by varying the pump intensity, and we track how the exciton resonance and sidebands are enhanced %
when increasing the exciton population.
We find that the exciton peak and the sidebands follow the same power law $N_\mathrm{ex}^{0.92}$ (see dashed lines in Fig.~\ref{fig:2}~(d)), confirming their common origin. This scaling deviates from the expected linear scaling, indicating that other processes such as probe-induced exciton dissociation are taking place during the probe pulse. We note that at high degree of exciton preparation, $\pm4\omega$ sidebands start to emerge in the HHG spectra but the associated enhancement is too low to be properly analyzed. Importantly, the direct relation between the population of bound excitons and the spectral weight of the exciton peak and sidebands opens the doors to ultrafast all-optical method of probing of exciton population. We also note that longer wavelengths seem to produce more intense exciton peaks and sidebands. It is therefore interesting to employ longer wavelengths to probe excitonic signatures.

In summary, we investigated how a prepared population of strongly bound excitons affect HHG in a one-dimensional solid, by modelling a pump-probe setup, thus getting insight into ultrafast exciton dynamics and revealed effects due to exciton dissociation and recombination. 
The behavior of exciton preparation under an intense pump was studied, revealing that the pump can dissociate the excitons it is creating, leading to more free carriers than excitons when the pump reaches high intensities.
The spectral features were found to be two fold: the presence of an exciton level allow for new excitation and recombination pathways. To reveal this, we developed an exciton-extended semiclassical interband model, from which we could explain how the carriers recombine into bound excitons. In addition, we observed how the presence of a finite population of bound excitons is able to enhance the HHG process in the region of the excitonic resonance, in particular leading to excitonic sidebands whose intensity was found to be close to proportional to the exciton population. 
We finally proposed HHG spectroscopy as a viable method of extracting information regarding a finite exciton population and exciton processes in solids exhibiting strongly bound excitons. The emergence of sensitive on-chip techniques for PHz-scale optical-field sampling provide experimental possibilities to temporally unravel such ultrafast light-driven exciton dynamics  \cite{PhysRevB.107.054302}.
There are still interesting  questions to be addressed. While we have investigated strongly-bound excitons, the role of continuum excitons as observed in semiconductors, carrier-induced screening, exciton-exciton interaction and dimensionality remains to be explored.

\begin{acknowledgments}
This work was supported by the Independent Research Fund Denmark (GrantNo.9040-00001B). S.V.B.J. further acknowledges support from the Danish Ministry of Higher Education and Science. Fruitful discussions with S. Latini is acknowledged.
This work was supported by the Cluster of Excellence Advanced Imaging of Matter (AIM), Grupos
Consolidados (IT1249-19), SFB925, “Light induced dynamics and control of correlated quantum systems”. The Flatiron Institute is a division of the Simons Foundation.

\end{acknowledgments}


\begin{thebibliography}{94}
\bibitem [{SM()}]{SM}%
  \BibitemOpen
  \href@noop {} {\bibinfo  {journal} {See {{S}}upplemental {{M}}aterial, which
  includes Refs. [67,69-83] for details on (I) numerical methods, (II) exciton
  model within the effective mass approximation, (III) HHG spectra for weakly
  excited or unexcited systems, (IV) exciton-seeded interband enhancement, (V)
  exciton-extended semiclassical interband model, and (VI) exciton energy
  shift}\ }\BibitemShut {NoStop}%
\end{thebibliography}
\end{document}

\title{Supplemental Material \\ High-harmonic spectroscopy of strongly bound excitons in solids}

\author{Simon Vendelbo Bylling Jensen}
\affiliation{Department of Physics and Astronomy, Aarhus University, DK-8000 Aarhus C, Denmark}
    
\author{Lars Bojer Madsen}
\affiliation{Department of Physics and Astronomy, Aarhus University, DK-8000 Aarhus C, Denmark}

\author{Angel Rubio}
\affiliation{Max Planck Institute for the Structure and Dynamics of Matter and Center for Free-Electron Laser Science, Hamburg 22761, Germany}
\affiliation{Center for Computational Quantum Physics (CCQ), The Flatiron Institute, New York, New York 10010, USA}

\author{Nicolas Tancogne-Dejean}
\affiliation{Max Planck Institute for the Structure and Dynamics of Matter and Center for Free-Electron Laser Science, Hamburg 22761, Germany}

\maketitle

\appendix
\section{I. Numerical Methods} \label{sec:app1}
\subsection{Time-dependent Hartree-Fock (TDHF) simulations}
For the TDHF simulations, we employ a hydrogen crystal system, which is known to exhibit strongly bound excitons \cite{PhysRevA.98.023415,PhysRevLett.127.077401}. Electrons are driven along the laser polarization direction through a periodic chain of hydrogen dimers with a bond length of $ 1.6 $ Bohr and lattice constant of $ a=3.6 $ Bohr. The interaction between nuclei, located at $ x_i $, and electrons are described with the softened Coulomb potential $v_{\mathrm{ion}} \left( x\right) = - \sum_i [\left(x-x_i \right)^2 + 1]^{-1/2}$. The dynamics are solved by propagating a set of orthonormal electron orbitals $\varphi_{i}^{\mathrm{HF}} \left(x ,t \right)$, through the velocity gauge formalism of the TDHF equation
\begin{equation}
i \partial_t \varphi_{i}^{\mathrm{HF}}(t) = \left\lbrace \frac{1}{2}\left[-i\partial_x +  A\left(t \right) \right]^2   + v_{\mathrm{ion}}(x) + v_H \left[n \right](x,t) + \widehat{v}_X \left[\lbrace \varphi_{j}^{\mathrm{HF}} (t)\rbrace\right] \right\rbrace \varphi_{i}^{\mathrm{HF}}(t).
\end{equation} 
Here $A\left(t\right)$ is the vector potential describing the applied laser pulse in the electric dipole approximation. The electron-electron interaction is described by the Hartree potential $v_H \left[ n \right] \left( x,t \right) = \int dx' n\left( x',t\right)[\left(x-x' \right)^2 + 1]^{-1/2}$, and the nonlocal exchange operator, which applied to an orbital is given as 
\begin{equation}
\widehat{v}_X \left[\lbrace \varphi_{j}^{\mathrm{HF}}(t) \rbrace\right]\varphi_{i}^{\mathrm{HF}}(t) = - \sum_{k=1}^{N} \varphi_{k}^{\mathrm{HF}}(x,t) \int dx' \varphi_{k}^{\mathrm{HF}*}(x',t)\varphi_{i}^{\mathrm{HF}}(x',t)[\left(x-x' \right)^2 + 1]^{-1/2}.
\end{equation}
 The density is $n \left(x,t\right) = \sum_{i=1}^{N} \abs{\varphi_{i}^{\rm HF} \left( x ,t\right)}^2 $. 
The ground state orbitals are obtained by a self-consistent iterative process of solving the HF equation starting from a linear combination of atomic orbitals using the Octopus software package \cite{doi:10.1063/1.5142502}. Hereafter, the TDHF equation is propagated using an enforced time-reversal symmetry propagator utilizing the adaptively compressed exchange operator formulation \cite{Lin2016}, and using a predictor-corrector scheme to guarantee that we reach self-consistency at every time step $\Delta t$, with $\Delta t=0.242$ as. Furthermore, we apply a singularity correction to treat divergent terms from the exchange energy \cite{PhysRevB.75.205126}, that we adapted to the one-dimensional case. %

\begin{figure}
\includegraphics[width=8.6 cm]{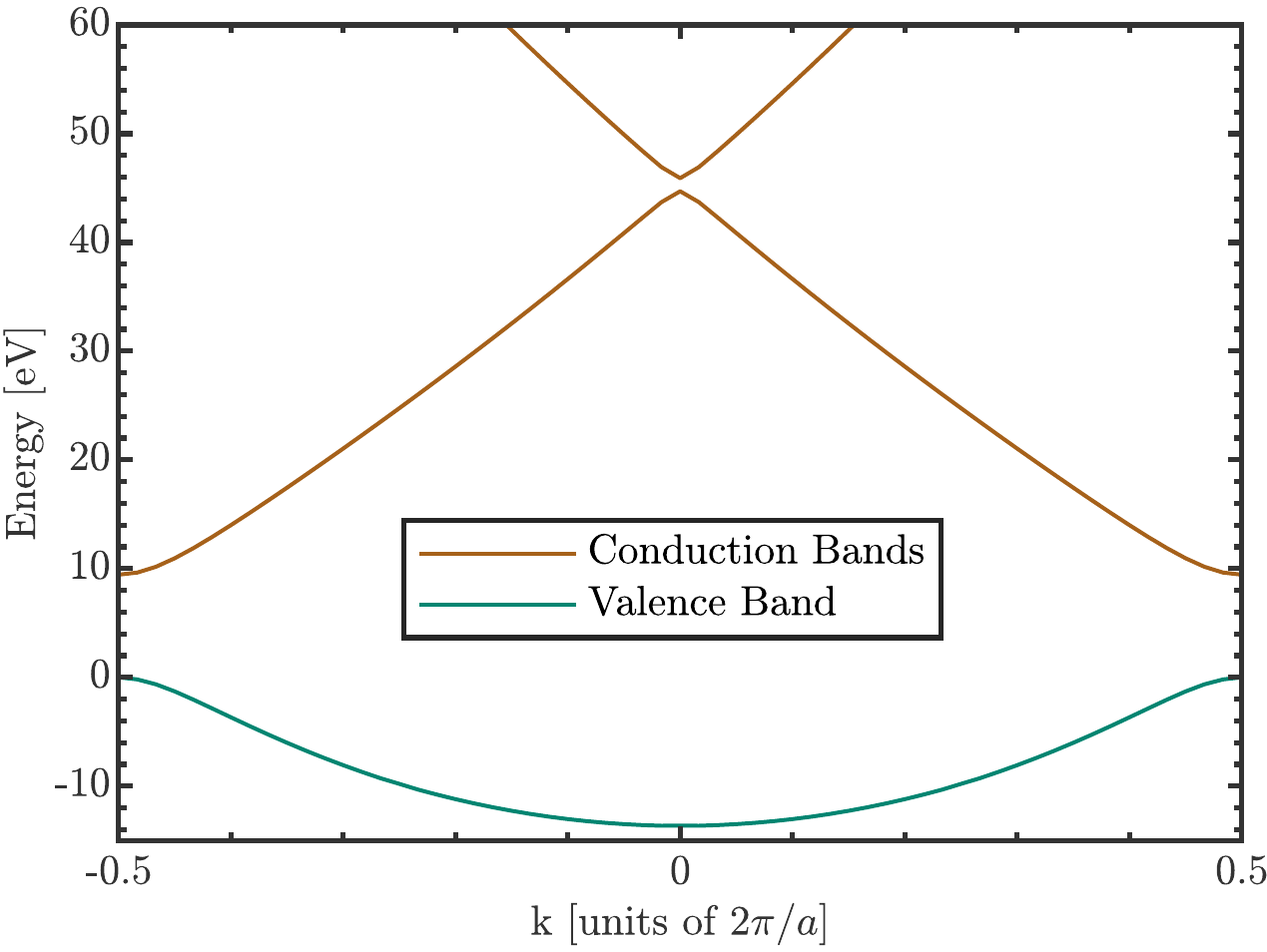}
\caption{Part of the band structure with the valence band and lowest energy conduction bands. The bandgap is found to be $9.45$ eV.} \label{fig:12}
\end{figure}

\begin{figure}
\includegraphics[width=8.6 cm]{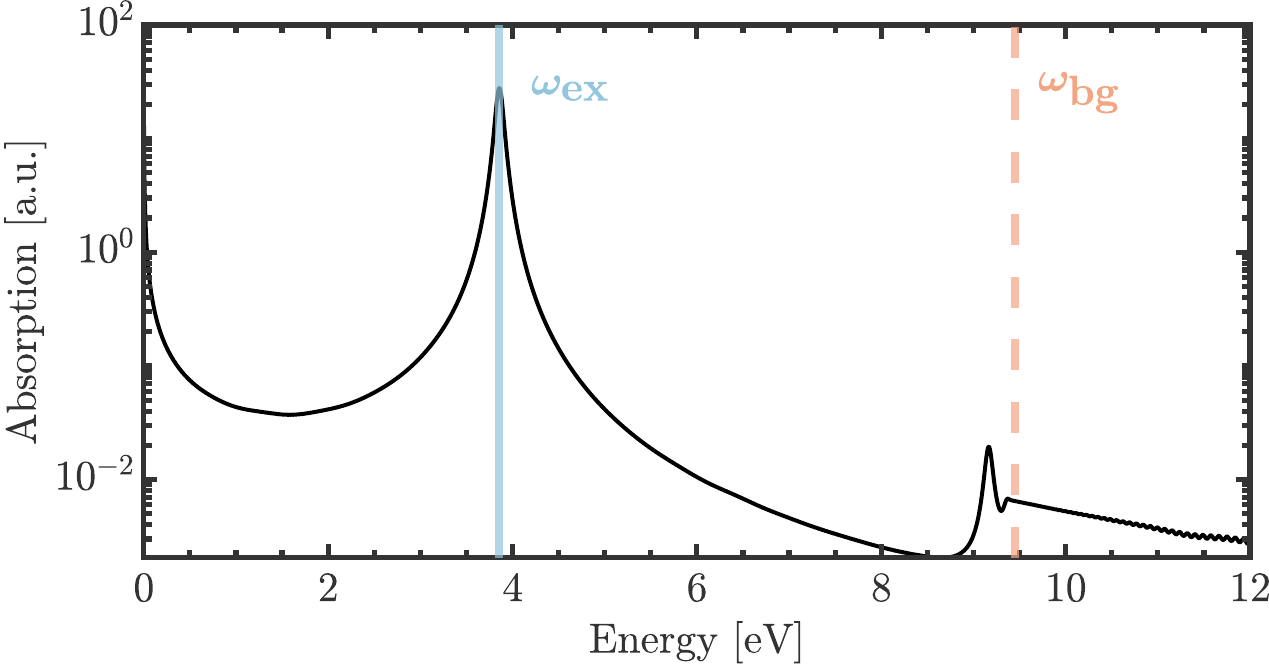}
\caption{Linear absorption spectrum calculated through TDHF with exponential dampening of the electronic current, corresponding to a\Srem{n} Lorentzian broadening of the absorption spectrum. The vertical lines denote the two applied pump frequencies, the exciton-resonant one, at $\omega_{\mathrm{ex}} = 3.86$ eV, and the bandgap-resonant one, at $\omega_{\mathrm{bg}} = 9.45$ eV. We associate the peak just below the bandgap energy to be a signature of excited excitonic states.} \label{fig:13}
\end{figure}
 The calculation for the periodic hydrogen chain model is performed using a converged real-space grid spacing of 0.4 Bohr. For the crystal momentum-space grid, convergence was achieved using 1024 grid points or increasing to 2048 grid points for the highly excited systems. Part of the bandstructure of the system is given in Fig.~\ref{fig:12}. From the band structure, we observe a direct bandgap of $9.45$ eV, which corresponds to $15.24$ harmonic orders of the HHG-generating probe pulse with a wavelength of $2000$ nm ($\sim 0.62$ eV). 
 
The linear absorption spectrum is obtained by the application of the small $\delta$-kick \cite{PhysRevB.62.7998} followed by a time evolution, and is shown in Fig.~\ref{fig:13}. With this, we obtain the pump frequencies for exciting the system. The first one,  $\omega_{\mathrm{ex}} = 3.86$ eV, corresponds to resonantly driving the transition to the strongly bound exciton. The second one, $\omega_{\mathrm{bg}} = 9.45$ eV, is resonant with the minimum bandgap, where valence and conduction bands have a high density of states. This second transition can provide energy for an exciton resonant transition from lower\Sadd{-}energy electrons of the valence band, however, with a weak coupling due to the low density of states at the associated section of the valence band.

\subsection{Time-resolved exciton wavefunctions}

The exciton wavefunction is constructed from the single-particle transition density matrix (TDM), by the procedure of Ref.~\cite{doi:10.1021/acs.jctc.0c01334}. The protocol relies on the transition density matrix, which has been helpful for the analysis and interpretation of excited states of molecular systems. When extended beyond the linear response regime for TDDFT and TDHF, the time-dependent TDM is constructed as a sum of weighted transition amplitudes. What makes it particularly useful, is that the TDM offers a way to resolve a given excitation in a spatial map constructed of pure single-particle excitations. In that way, the time-dependent TDM\Sadd{,} $\Gamma_s (x,x',t)$\Sadd{,} for a given excitation represents weighted transition amplitudes of processes where a particle is annihilated at position $x$ and created at position $x'$ \cite{LI2011157}. One can thus think of the TDM as accounting for the conditional probability of finding an electron at a certain place given the position of the associated hole. 
For real-time TDHF, the time-dependent TDM can be calculated at a specific time $t$ utilizing the HF orbitals at that time $\varphi_{\sigma,i}^{\mathrm{HF}} \left(x ,t \right)$, which has been propagated from their ground state $\varphi_{\sigma,i}^{\mathrm{HF}} \left(x,t=0\right)=\varphi_{\sigma,i}^{\mathrm{HF}} \left(x\right)$. The time-dependent TDM is denoted here as the difference between the time-dependent and the ground state one-body density matrices, which can be evaluated as
\begin{equation}
\Gamma_s (x, x', t) = \sum_i^{\text{occ}}\left[\varphi_{i}^{\mathrm{HF}} \left(x ,t \right) \varphi_{i}^{\mathrm{HF}\ast} \left(x' ,t \right) - \varphi_{i}^{\mathrm{HF}} \left(x\right) \varphi_{i}^{\mathrm{HF}\ast}\left(x' \right) \right]. \nonumber 
\end{equation}
For periodic systems, it is expressed using the knowledge of orbitals inside the unit cell thanks to Bloch's theorem. However, the distances $x$ and $x'$ of the time-dependent TDM reside in the full crystal, given by the unit cell volume multiplied by the number of $k$-points used in the simulation. Indeed, the exciton occupies the entire crystal structure. To associate some physical insight into the TDM, we note that the diagonal of the TDM contains the time-dependent density response associated with the given excitation $\Gamma_s (x, x, t) = \delta n(x,t)$, with $\delta n(x,t) = n(x,t) - n(x,t=0)$. Since the integrated density remains constant with time, this enforces that $\int dx \left[ \Gamma_s (x,x,t) \right] = 0$. In quantum chemistry, the time-dependent TDM has been shown to be analogous to the exciton wavefunction given from Green's function theory \cite{C5CP07077E}. In obtaining the time-dependent TDM in real space, one can extract information with regards to the dynamical exciton processes, such as dissociation rates or charge separation rates. In the present work, we fix the hole at $x'=0$. A complete quantitative analysis would require to scan both the electron and hole coordinate. But due to the large-scale structure of the TDM being diagonally dominated, we conclude that qualitative features can be inferred from the TDM with the hole placed at the center between the two nuclei. In the following and in the main text, we refer to $\Gamma_s (x, x'=0, t)$ as the exciton wavefunction for the case of the hole being localized at $x'=0$.

\subsection{First moment of the exciton wave function}
In the reference frame of the hole, the first moment is given as 
\begin{equation}
m = \frac{\int dx \abs{x} \abs{\Gamma_s (x,0,t)}^2}{\int dx \abs{\Gamma_s (x,0,t)}^2}
\end{equation}
and normalized to the excitation magnitude. As observed in Fig.~1 of the main text and argued in the main text, the system prepared with bound excitons have a first moment of a few \AA. If the system is pumped with free carriers, then the first moment will be hundreds of \AA~for our system.

\subsection{Number of excitons and free carriers}

We now want to extract information from the exciton wavefunction $\Gamma_s (x, x'=0, t)$, to give a description of the number of excitons or free carriers in a given excitation. To do this, we consider the exciton wavefunction and assume that it consists of the contributions from excitons and free carriers. We assume that exciton population, gives a localized exciton wavefunction, or in other words, a large conditional probability of finding the electron near its hole. As observed in Figs.~1~(c),(d) of the main text, this contribution has an exponential scaling with distance and a first moment of less than $5$ \AA. The other contribution to the exciton wavefunction is originating from free carriers. We assume that these latter carriers provide a relatively uniform distribution of the exciton wavefunction with regards to distance. This is since the excited carriers are traversing freely throughout the lattice and thus have a uniform conditional probability for finding electron relative to its hole. Based on these two different contributions to the exciton wavefunction, we can interpret the ratio of such contributions to imply the ratio of excitons to free carriers for a given excitation. Normalizing these measures with the number of excited valence electrons per unit cell, $N_\text{e}$, we can obtain the number of excitons $N_\text{ex}$ and the number of free carriers $N_\text{fc}$ for a given excitation. 

The number of free carriers can be calculated as the contribution of the exciton wavefunction, which is uniform with respect to distance. To calculate this contribution, we integrate the exciton wavefunction from beyond the point at which the excitonic contribution is dominant, which we denote as the exciton radius $x_{\text{ex}}$. The average value of the exciton wavefunction in this region is then extended across the full crystal length $L$ to give the free-carrier contribution to the exciton wavefunction. The ratio of this contribution to the total size of the exciton wavefunction, provide the ratio of free carriers for the excitation and multiplying with number of excited valence electrons, we obtain the number of free carriers as

\begin{equation}
N_\text{fc}(t) = N_\text{e} (t)  \frac{L}{L-x_{\text{ex}}} \frac{\int^{L}_{x_{\text{ex}}} dx  |\Gamma_s (x,0,t)|^2 }{ \int^{L}_0 dx | \Gamma_s (x,0,t)|^2}. \label{eq:fc}
\end{equation}

From Fig.~1 (a) in the main text we identify $r_{\text{ex}} = 5$ Å as the radius of the bound exciton, in the low-exciton concentration regime, and we use this $x_{\text{ex}}$ for characterising the excitonic localization. The number of excited valence electrons per unit cell is defined as
\begin{equation}
N_\text{e} (t) =  N_\text{tot}  - \sum^\text{occ}_{n,n'}  \abs{\bra{\varphi_{n}^\mathrm{HF} (t)} \ket{\varphi_{n'}^\mathrm{HF} (t=0)} }^2 \label{eq:e}
\end{equation}
and is computed during our simulations, with $N_\text{tot}$ being the total number of valence electrons. This allows us to compute $N_\text{fc}(t)$ from $N_\text{e} (t)$ and $\Gamma_s (x,0,t)$. The remaining excitation must then consist of excitons, which can be found from 
\begin{equation}
N_\text{ex} (t) =  N_\text{e} (t) - N_\text{fc} (t)  \label{eq:ex} 
\end{equation}
We note, that with the definitions from Eqs.~\eqref{eq:fc},\eqref{eq:e} and \eqref{eq:ex} a completely uniform exciton wavefunction will give $N_\text{fc} = N_\text{e}$ and $N_\text{ex}=0$. Similarly the limit of completely localized electron wavefunction within $r_\text{ex}$ will provide $N_\text{fc} = 0 $ and $N_\text{ex} = N_\text{e}$. 

\section{II. Exciton model within the effective mass approximation}

To further understand the effect of the probe laser on the exciton population, we modelled the dynamics of the exciton within the effect mass approximation.
For this we follow the works of Ogawa and Takagahara~\cite{PhysRevB.43.14325,PhysRevB.44.8138} that we extend to the time-dependent case. Here we review the derivation starting from a three-dimensional two-body electron-hole time-dependent Schr\"odinger equation within a laser field linearly polarized along the $z$ direction 
\begin{eqnarray}
 i\hbar\partial_t \Phi(\mathbf{r_e},\mathbf{r_h},t) = \Big(\frac{1}{2m_e}(-i\hbar\nabla_e-\frac{e}{c}\mathbf{A}(t))^2 +\frac{1}{2m_h}(-i\hbar\nabla_h+\frac{e}{c}\mathbf{A}(t))^2 + U_e(\mathbf{r_e}) + \nonumber\\ U_h(\mathbf{r_h}) + V(\mathbf{r_e},\mathbf{r_h}) \Big) \Phi(\mathbf{r_e},\mathbf{r_h},t)\,,\nonumber
\end{eqnarray}
where $\Phi(\mathbf{r_e},\mathbf{r_h})$ is commonly referred to as the exciton envelope function, $m_e$ ($m_h$) is the effective mass of an electron (a hole), $V$ is the Coulomb interaction, $U_e$ ($U_h$) the confining potential acting on the electron (hole), and $\mathbf{A}(t)=A(t)\mathbf{\hat{e}_z}$ is the vector potential of the laser.\\

Assuming a strong confinement of the carriers in the lateral directions, we can apply the envelope approximation, which assumes that~\cite{PhysRevB.44.8138}
\begin{equation}
 \Phi(\mathbf{r_e},\mathbf{r_h},t) = e^{iKZ}f_e(x_e,y_e,t)f_h(x_h,y_h,t)\phi(\mathbf{r_e}-\mathbf{r_h},t)\,,
\end{equation}
where $Z$ and $K$ are  $z$ coordinate of the center of mass of the exciton and the corresponding wavenumber, and $f_e$ and $f_h$ are the lowest sub-band functions in the lateral directions for the electron and the hole. Finally $\phi$ describes the relative motion of the electron and the hole. 
After some algebra, and assuming that $f_e$ and $f_h$ remain normalized to unity at all times and that $\phi$ only depend on $z=z_e-z_h$, we obtain a one-dimensional time-dependent Schr\"odinger equation
\begin{eqnarray}
 i\hbar  \partial_t \phi(z,t) = \Big(\frac{1}{2\mu}(\partial_z^2 -\frac{e}{c}A(t))+ \frac{\hbar^2}{2(m_e+m_h)}K^2
 + V_{\rm eff}(z) + eE(t)z  \Big)  \phi(z,t) \,,\nonumber
\end{eqnarray}
where $\mu$ is the exciton reduced mass, $V_{\rm eff}(z) = \int dx_edy_edx_hdy_h V(\mathbf{r_e},\mathbf{r_h}) |f_e(x_e,y_e)|^2|f_h(x_e,y_e)|^2$ is the confinement potential. We used here the fact that the confining potentials $U_e$ and $U_h$ are time independent, and hence the $V_{\rm eff}$ is time independent. Assuming that the energy of the center-of-mass motion $\frac{\hbar^2}{2(m_e+m_h)}K^2$ remain constant during the time evolution, this term can be transformed away and we have thus reduced the exciton problem to time-evolution of a one-dimensional problem of a single particle within an electric field, with the only change that the particle has a reduced mass $\mu$.
This equation is easily solved using any software capable of propagating in time a one-dimension time-dependent Scr\"odinger equation, and in this work, this is numerically solved using the Octopus code~\cite{doi:10.1063/1.5142502}.\\

\begin{figure}[t]
  \begin{center}
    \includegraphics[width=0.6\columnwidth]{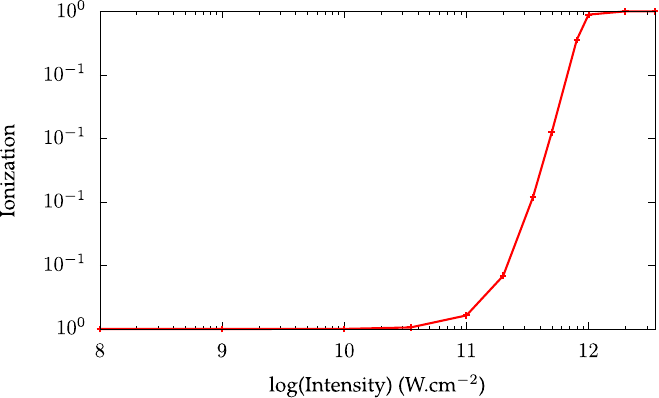}
    \caption{Ionization of the exciton induced by a pump laser as in the main text versus different intensities. We employ the same laser parameters as for Fig. 1 in the main text.}\label{fig:populations}
  \end{center}
\end{figure}

In order to estimate how much the exciton are dissociated by the pump itself, we now perform simulations of the exciton dynamics under the envelope approximation, as described above, to simulate the dissociation process induced by pump laser. For this, we let the exciton wavepacket evolve in time under the influence of the pump laser and we place absorbing boundaries at a distance of 500 Bohr from the center of the simulation box, to absorb the part of the wavepacket that dissociates under the influence of the laser.
This allows us to estimate the fraction of dissociated exciton population for various laser intensities. The results of the simulations are shown in Fig.~\ref{fig:populations}. We employed here a grid spacing of 0.4 Bohr, and used a simulation for a radius 600 Bohr, including 100 Bohr wide absorbing boundary region using a complex absorbing potential of height -0.2 with a sin-square envelope~\cite{de2015modeling}. \\

This simple modelling shows that pump-induced exciton dissociation is a very important when the intensity reaches values closes to 10$^{11}$ W cm$^{-2}$ and above.

\section{III. HHG spectra for weakly excited or unexcited systems}
\begin{figure} 
\includegraphics[width=8.6 cm]{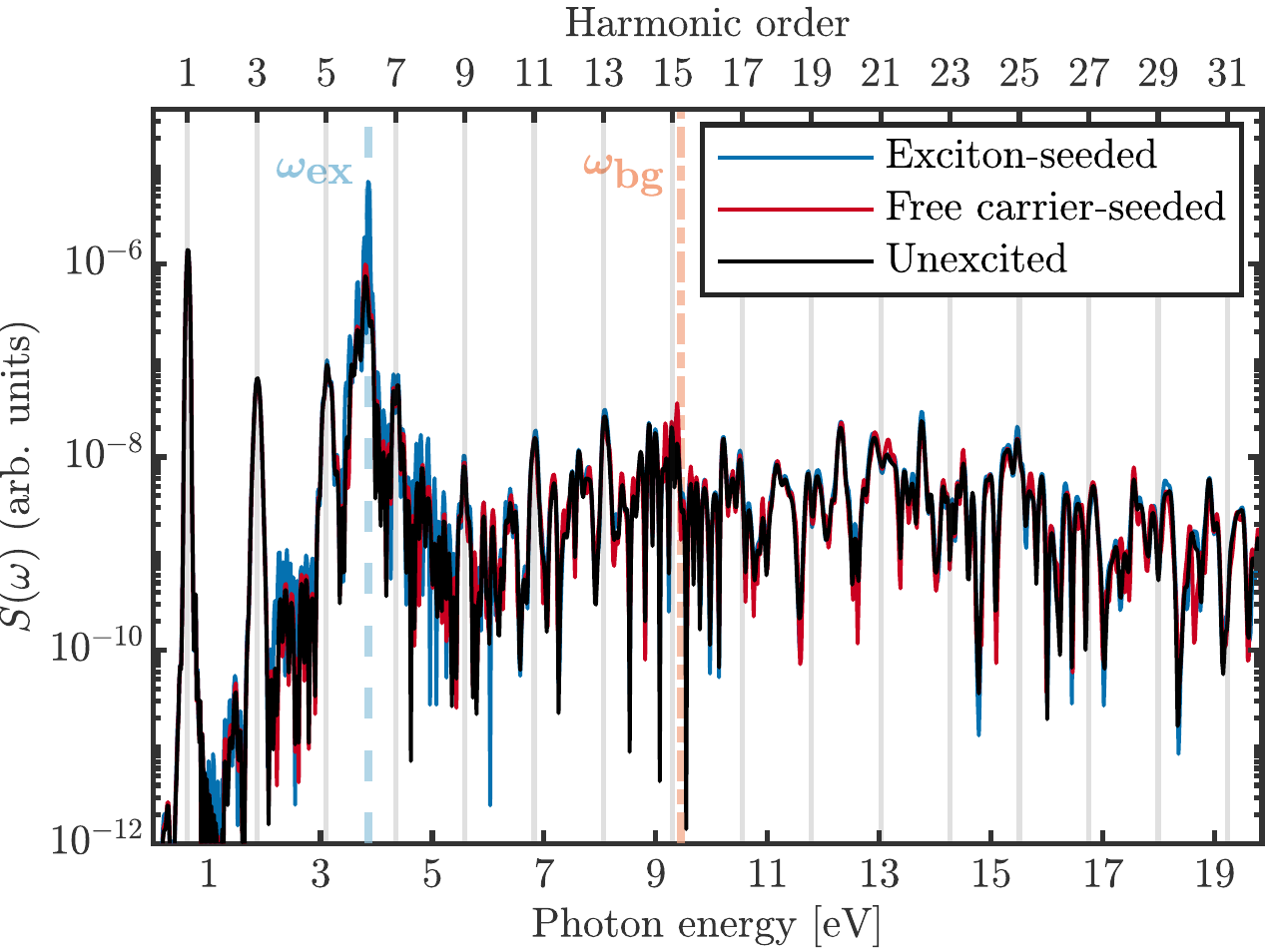}
\caption{High-harmonic generation spectra for systems that are seeded to an insignificant degree with excitons and free carriers and compared to the spectrum for an unexcited system. The excited systems are prepared with a $10^6$ W cm$^{-2}$ $\omega_{\mathrm{ex}}$-pump or a $10^{10}$ W cm$^{-2}$ $\omega_{\mathrm{bg}}$-pump to generate a weak excitation of, respectively, bound excitons or free carriers as given in Fig.~1 (f) of the main text. All system are driven with a $2000$ nm probe of intensity $10^{12}$ W cm$^{-2}$. The exciton resonance and the bandgap energy is marked with a blue dashed and red dashed-dotted line, respectively. See text in Sec. III for pulse durations and delay.} \label{fig:7}
\end{figure}

Throughout our work a $\sin^2$-pulseshape is applied for both pump and probe pulses. The duration of the pump (probe) pulse is $25$ fs ($100$ fs) with a peak-to-peak delay of $72.5$ fs. The probe intensity is kept at $10^{12}$ W cm$^{-2}$, whereas a wide scan of pump intensities has been explored. For the pump-probe systems, we consider an experiment of orthogonally oriented polarization for the pump and probe pulse respectively, such that the perturbative response of the pump is not present in the HHG spectra. To account for this along the one-dimensional model, we only consider the probe-induced current for the spectra, which is the current induced by the probe when affecting a pump-prepared system. Numerically this corresponds to considering $J(t) = J_{\text{pump-probe}}(t) - J_{\text{pump}}(t)$, with $J_{\text{pump-probe}}(t)$ being the system response to a pump-probe scheme, and $J_{\text{pump}}(t)$ being the system response to only a pump pulse. Here $J(t)$ is the electric current, evaluated by 
\begin{equation}
J(t)= \sum_{i,\sigma} \int dx \Re \left[\varphi^{\text{HF}\ast} _{i,\sigma} \left( x,t \right) \left( -i \pdv{x} + A\left(t\right)\right) \varphi_{i,\sigma}^{\text{HF}} \left( x,t \right)  \right].
\end{equation}
From this, the HHG spectra are obtained as 
\begin{equation}
S(\omega) \propto  \abs{ \omega \int dt J\left( t \right) e^{-i\omega t}}^2
\end{equation}
When performing the Fourier transform, a window function is applied with a $\cos^8$ decay to attenuate the current at the very end of the simulated interval to avoid numerical artifacts.
We have investigated the response for samples prepared with a low degree of excitation of either excitons or free carriers in Fig.~\ref{fig:7}. The exciton-seeded system is prepared with a $10^6$ W cm$^{-2}$ $\omega_{\mathrm{ex}}$-pump and the free carrier-seeded system is prepared with a $10^{10}$ W cm$^{-2}$ $\omega_{\mathrm{bg}}$-pump. The degree of excitation can be retrieved from Fig.~1 (f) in the main text. We observe the harmonic emission of both seeded systems to largely resemble the response from the unpumped system. The small discrepancies are observed around the pump-frequencies. An intensity scan reveals, that for modification of the HHG process due to the seeding of the sample, the pump must provide an excitation of $10^{-5}-10^{-4}$ valence electrons per unit cell. For the unpumped system, we recover similar features as in Ref.~\cite{PhysRevA.98.023415}.

\section{IV. Exciton-seeded interband enhancement}
\begin{figure} 
\includegraphics[width=8.6 cm]{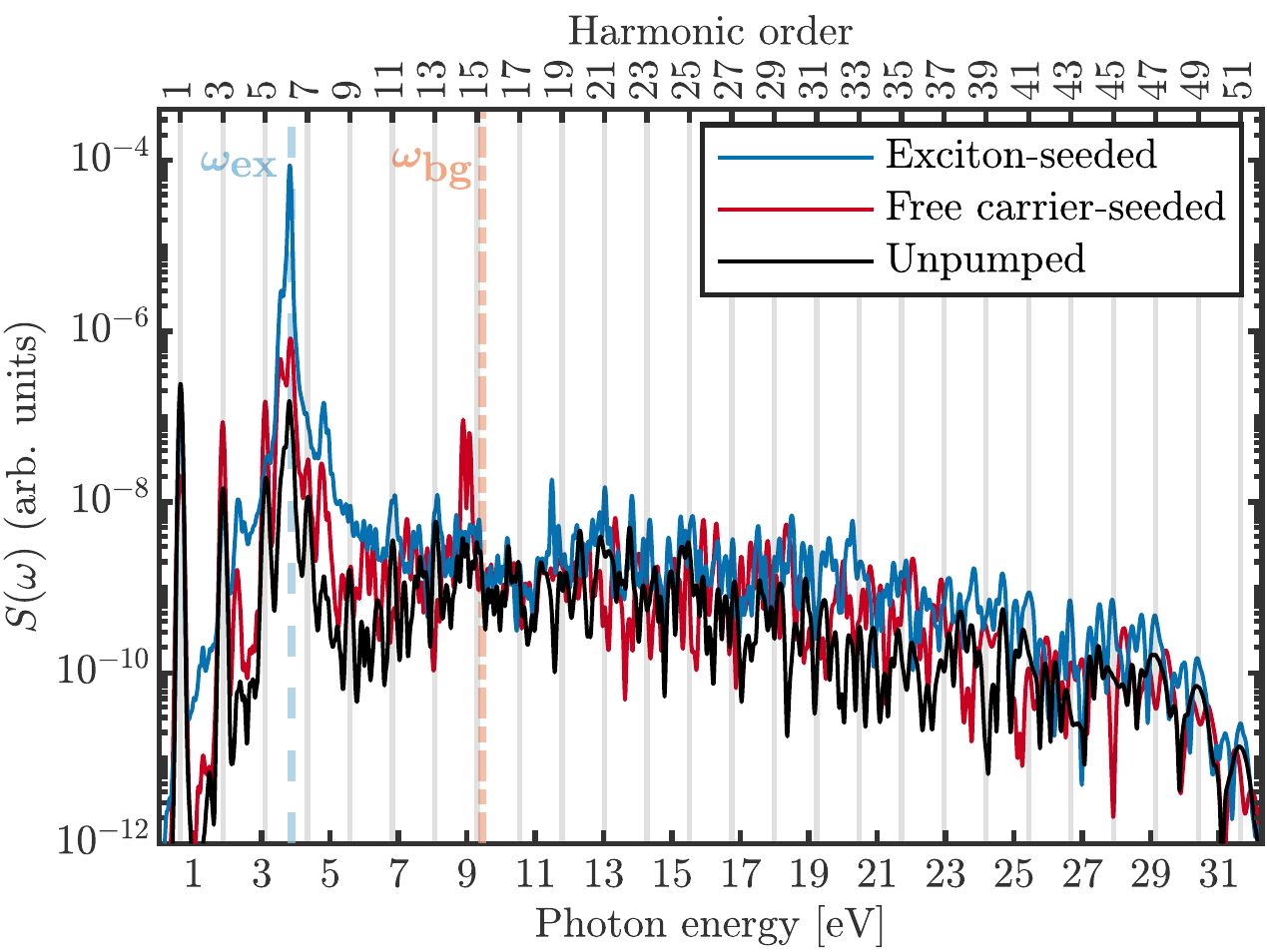}
\caption{High-harmonic generation spectra for the systems prepared with a large population of bound excitons with $\omega_\mathrm{ex}$ or free carriers with $\omega_\mathrm{bg}$. The excited systems are prepared with a $10^8$ W cm$^{-2}$ $\omega_{\mathrm{ex}}$-pump or a $10^{12}$ W cm$^{-2}$ $\omega_{\mathrm{bg}}$-pump to generate a strong excitation of, respectively, bound excitons or free carriers. The total excitation prepared by the $\omega_\mathrm{bg}$-pump is $2.3$ times larger than the excitation generated by the $\omega_\mathrm{ex}$-pump, as given in Fig.~1 (f) of the main text. The harmonics are obtained when driven with a $2000$ nm probe of intensity $10^{12}$ W cm$^{-2}$. The exciton resonance and the bandgap energy is marked with a blue dashed and red dashed-dotted line, respectively. For illustrative purposes, the spectra have been smoothed. See text in Sec. III for pulse durations and delay.} \label{fig:8}
\end{figure}
\begin{figure} 
\includegraphics[width=8.6 cm]{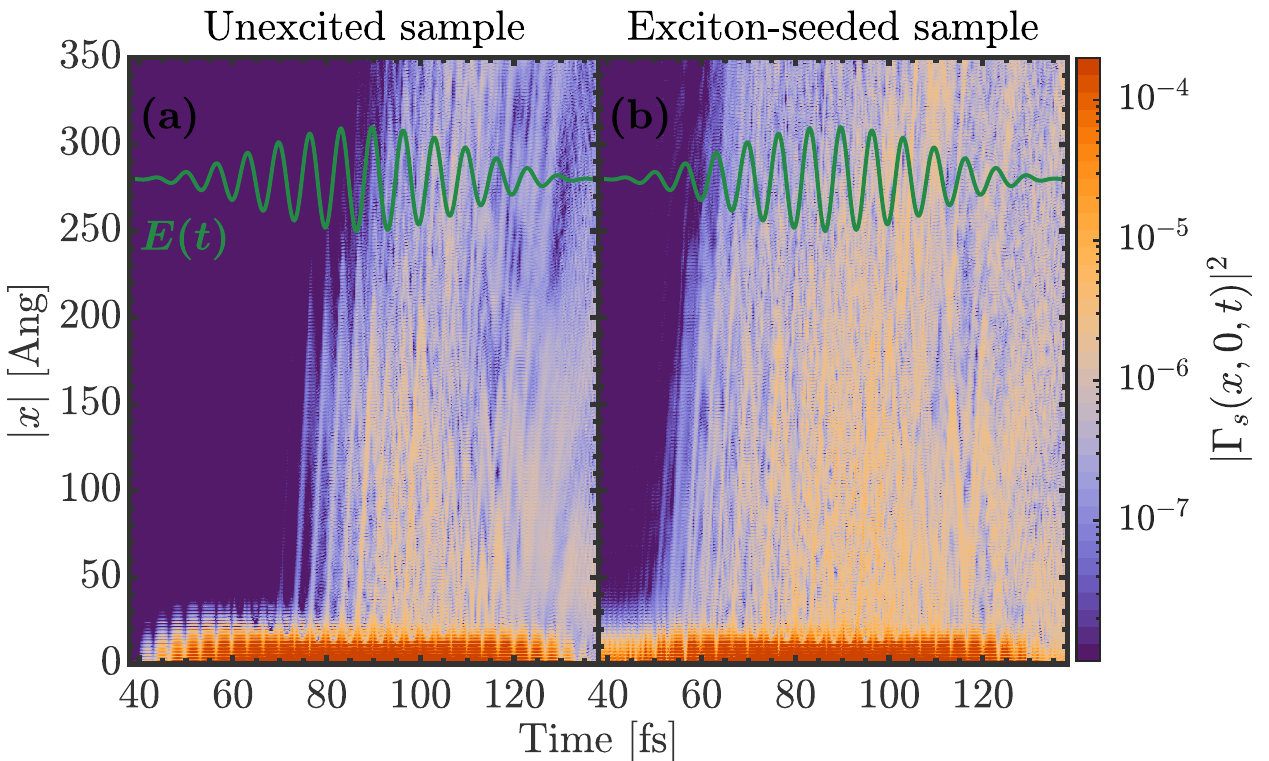}
\caption{Norm squared exciton wavefunction as a function of distance and time, during the HHG process. Comparing an (a) unpumped system with (b) an exciton-seeded system with parameters of Fig.~\ref{fig:8}. The exciton wavefunction is considered within the temporal region of the probe pulse, which electric field is inserted in green.} \label{fig:9}
\end{figure}

In the strong excitation regime, we have investigated the effect of a large degree of free carriers or exciton excitation. To do this, we consider a bound exciton-seeded system prepared with a $10^8$ W cm$^{-2}$ $\omega_{\mathrm{ex}}$-pump, which is known to produce a large population of bound excitons from Fig.~1 (f) in the main text. To compare, we chose a $10^{12}$ W cm$^{-2}$ $\omega_{\mathrm{bg}}$-pump, as this generates an excitation, which is $2.3$ times larger in magnitude than the $\omega_{\mathrm{ex}}$-pump, but consist mainly of free carriers, but also a fraction~1\% of bound excitons. The excitons can be generated both as a consequence of free carriers recombining to form excitons, or by the $\omega_\mathrm{bg}$-pump coupling directly to excited exciton states from the valence band. Both system are now prepared in a state with a large excitation and we thus observe a more convoluted system response in Fig.~\ref{fig:8}. Considering, e.g., the free carrier-seeded system response, we see an enhancement of the exciton resonance and sidepeaks as well, since this system is also seeded with an significant fraction of excitons. Actually, the number of excitons generated here is comparable with the exciton-seeded system of Fig.~3 (b) in the main text, which show a similar enhancement of the exciton-related peaks. For the intense exciton-seeded spectrum in Fig.~\ref{fig:8}, we observe a more significant enhancement of the exciton peak and sidepeaks. Furthermore, here we also observe the exciton-seeded sample to have an enhancement across the first plateau. This enhancement can be explained as a significant part of the seeded bound exciton population can dissociate to free carriers, during the probe pulse. This dissociation is directly observed if considering the exciton wavefunction in Fig.~\ref{fig:9} (b) where an additional contribution is observed from dissociated excitons, which form carriers into the first conduction band and can contribute to intraband transitions across the first plateau, as observed in Fig.~2 (b) in the main text.  Interestingly, our TDHF calculation predicts, that seeding the sample with excitons is expected to be more efficient than seeding the sample with free carriers for above bandgap harmonic enhancement. This is despite of the fact, that here in Fig.~\ref{fig:8}, degree of excitation in the free carrier-seeded system is $2.3$ times larger than the excitation of the exciton-seeded sample. In Fig.~\ref{fig:9} (a) we also observe, that around $75$ fs, a substantial dissociation of excitons is observed, which correspond to the time at which the interband harmonics are initially generated in Fig.~2~(a) in the main text.

\section{V. Exciton-extended semiclassical interband model}

For interband HHG processes a trajectory-based semiclassical model have been developed based on solving the semiconductor Bloch equations with the saddle-point approximation \cite{PhysRevB.91.064302}. It has succeeded in describing cutoff energies, and extended to capture imperfect collisions and Umklapp scattering \cite{PhysRevA.102.033105,PhysRevResearch.2.033333}. Here we extend this model to include recombination paths associated with excitons, but note that imperfect collisions and Umklapp scattering might also intricate the dynamics here. First step of the interband model, is the generation of a electron-hole pair at the bandgap of the solid. We do not distinguish whether the electron-hole pair is generated from exciting an electron from the valence to conduction band, or whether these are generated by the dissociation of an exciton. As both these processes will most likely result in the generation of a electron-hole pair at the bandgap of the solid. The relative distance between the electron and hole $\Delta x(t)$ is now propagated from the generation time $t_0$ using the conduction and valence band velocity $v_c(k)$ and $v_v(k)$ as
\begin{equation}
\Delta x(t)= \int_{t_0}^t\lbrace v_c[k(t')]-v_v[k(t')] \rbrace dt', 
\end{equation}
The relative velocity is expressed from the curvature of the bandstructure for the valence and conduction bands, $\varepsilon_{v}(k)$ and $\varepsilon_{c}(k)$ as
\begin{equation}
v_c(k(t'))-v_v(k(t')) = \pdv{\left[\varepsilon_{c}\left(k(t')\right) - \varepsilon_{v}\left(k(t') \right) \right]}{k}
\end{equation}
with crystal momenta governed by the acceleration theorem  
\begin{equation}
k(t)= k_0 + A(t).
\end{equation}
During propagation, the electron and hole might recombine at a recombination time $t_r$, where $\Delta x(t_r)=0$. At the recombination step, the electron and hole pair is assumed to recombine and emit their excess energy. This leads to two pathways, (1) recombination to the valence band, and (2) recombination to an exciton.  The second pathway is the exciton-extension to the model, and is constructed similar to the recombination into a donor-doped state of Ref. \cite{PhysRevA.99.013435}. The energies emitted for the different recombination steps are given as
 \begin{enumerate}
     \item Recombination with emission of the electron-hole pair energy of $\varepsilon_{c}\left[k\left(t_r\right)\right] - \varepsilon_{v}\left[k\left(t_r\right)\right]$
    \item Recombination into an exciton of exciton binding energy $E_b$ with emission of $\varepsilon_{c}\left[k\left(t_r\right)\right] - E_b$
 \end{enumerate}
With these two pathways of recombination, the time-frequency profile of the harmonic emission is two bands, following the well\Sadd{-}established interband emission chirp.

\section{VI. Exciton energy shift}

\begin{figure} 
\includegraphics[width=17.2 cm]{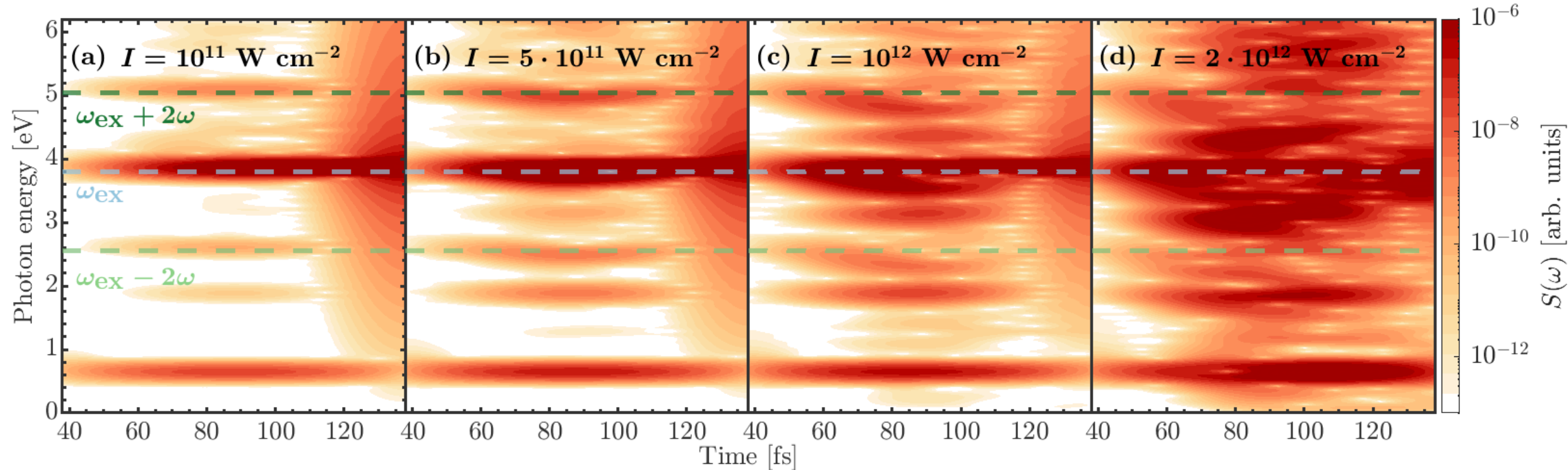}
\caption{Time-frequency analysis of the below-bandgap harmonic radiation for an exciton-seeded sample, obtained by an exciton-resonant pump with an intensity of $10^7$ W cm$^{-2}$. The probe pulse intensity is scanned in the regime of $10^{11}$ to $2 \times 10^{12}$ W cm$^{-1}$ and given in the caption. We used a window of  $\sigma = 10$ fs for the Gabor transform. The position of the exciton peak and the first exciton sidebands are denoted with dashed lines.} \label{fig:10}
\end{figure}

In the time\Sadd{-}frequency analysis of Fig. 2 of the main text, it is observed that the exciton features of the harmonic spectra is shifted in energy during the interaction with the driving pulse. This leads to\Sadd{,} e.g.\Sadd{,} the exciton features of the harmonic spectrum having a wider spectral width, when considering the HHG spectra of Fig. 3 of the main text. Such energy shift could be a result of a weakening of the binding energy due to presence of excited carriers, or mechanisms such as the excitonic Stark effect. We report the energy shift to be increasing with the intensity of the probe pulse, and thus the degree of excited carriers. To support this, we provide time\Sadd{-}frequency analysis with a scan of probe intensities in the regime of $10^{11}$ to $2 \times 10^{12}$ W cm$^{-1}$ for the exciton-seeded sample of Fig.~2~(d) in the main text. A time-frequency analysis of the emitted harmonics during the probe pulse is given in Fig. \ref{fig:10}, and we clearly observe how the exciton resonance, and sidebands have a larger displacement in energy at higher pulse intensities.